\documentclass[noshowpacs, twocolumn,notitlepage,preprintnumbers,secnumarabic,amsmath,superscriptaddress, showkeys, amssymb,nofootinbib,longbibliography]{revtex4-1}
\usepackage[compact]{titlesec}
\usepackage{bm}           
\usepackage{graphicx,bm}
\usepackage{caption}
\usepackage{gensymb}
\usepackage{float}
\usepackage[pdftex,bookmarks,colorlinks,breaklinks]{hyperref}  
\usepackage{mathrsfs}
\usepackage{color}
\usepackage[table]{xcolor}
\begin{document}
\title{Analogue Magnetism: An Ansatz}
\author{Bob Osano} 
\email{bob.osano@uct.ac.za} \affiliation{Cosmology and Gravity Group, Department of Mathematics and Applied Mathematics, and}
\affiliation{Academic Development Programme, Science Unit, University of Cape Town, Rondebosch 7701, Cape Town, South Africa}
\author{Patrick W. M. Adams}
\email{pw.adams@uct.ac.za} \affiliation{Cosmology and Gravity Group, Department of Mathematics and Applied Mathematics, and}

\begin{abstract} Similarities between magnetic flux density and fluid vorticity equations for the cases of negligible diffusion and viscosity are often considered an indication of the possible existence of physical analogue between the two. In this letter, we extend the comparison to cases where neither diffusion nor viscosity are negligible. An ansatz that relates the two fluids is established and the case of vanishing diffusion and viscosity is shown to be sub-case of the general. We propose that the magnetic flux density evolution equation be compared to the evolution equation of an effective vorticity ($\omega_{eff}$); related to the ordinary vorticity via a power law.
\end{abstract}

\date{\today}

\maketitle

\section{Introduction}
The analogy between fluid dynamics and electromagnetism has a long history and dates to when Maxwell\cite{Max} laid down the foundation of electromagnetism. In this brief letter, we will recount the development of the subject, it suffices to say that analogies involving electromagnetism have been studied for low-to-high Reynolds numbers as illustrated in the (\cite{Bel} - \cite{Arb} and \cite{BP} -\cite{BP}, however a theory of {\it analogue magnetism} remains illusive. It is clear from these works that a broad range of methodologies have been applied in an effort to gain a better understanding of the the two fluids, and to compare fluids where diffusion and viscosity are negligible. These range from mathematical analytical methods to numerical simulations methods. Researchers in this area of study are split into two schools of thought. One explores and emphasizes {\it similarities} while the other the {\it differences}. Whether or not {\it similarities} are more important than {\it differences}, both schools of thought contribute significantly to our understanding of the behaviors of charged and uncharged fluids. We think that it is possible to examine the two fluids whilst asking under under what {\it parametric conditions} does the analogy hold or breakdown? This is not dissimilar to the dynamo vs anti-dynamo studies and debates that have taken place, and is the motivation for this brief letter. We first consider a recent example of this debate.

The author of \cite{Ohki} recently presented the findings of a study where nonlinear vortex stretching for incompressible Navier - Stokes turbulence was compared to a linear stretching process of passive vectors where focus was given to the difference rather than the similarities between these processes under long and short time evolutions. This was a followup to the earlier work in \cite{Ohki2} where it was found that the vortex stretching effect ( $\omega . {\bf S}.\omega$) of vorticity is weaker than a similar effect in a general passive vector (magnetic flux density is an example of a passive vector). Here ${\bf S}$ is the Savart constraint \cite{Ohki}. Although the tools employed in the study are statistical, the findings are suggestive of a general behavior of the magnetic flux density flow. The difference is explained at a more physically fundamental level via the Biot-Savart formula. Besides the vortex stretching another kind of stretching is already embedded in the inductive part of the evolution equation, and which resolves to ${\bf B}.\nabla{\bf v}$  for the magnetic equations or $\omega .\nabla{\bf u}$ for the fluid equation.  A hint of the explanation for this is given in \cite{BP} and is related to how Navier-Stokes (NS) equations are affected by the individual constituent terms. In particular, it is argued in \cite{BP} that the presence of pressure reduces the magnitude of the velocity field that emerges from the corresponding NS equation, whereas pressure increases the magnitude of velocity that emerges from the NS equation corresponding to the magnetic field ( or any other passive vectors field). This means that the two fields will experience different stretching. The question of whether this faults the much sought-after analogy is still open for debate. We here give a possible way forward.\\ \\ 
\section{Key Equations}
We keep the discussion simple by focussing on the two main equation, namely the magnetic flux density and the vorticity evolution equations: 
\begin{eqnarray}
\label{one1}\frac{\partial{\bold{{\bf B}}}}{\partial{t}}&=&\nabla\times(\bold{v}\times{{\bf B}})+\eta\nabla^{2}{{\bf B}},\\
\label{one2}\frac{\partial{\bold{\omega}}}{\partial{t}}&=&\nabla\times(\bold{u}\times{\omega})+\eta\nabla^{2}{\omega},
\end{eqnarray} 
where ${\bf B}$ is the magnetic flux density, $\omega$ is the fluid vorticity, $\eta$ is the magnetic diffusion and $\nu$ is the kinematic viscosity. Of course, these are linked to their NS or momentum equations wherein we only consider couplings that are devoid of the Lorentz force. The two equations are then numerically simulated as indicated below. \\ 
\section{Main Results}
\begin{figure}[H]

 	\centering
		\caption{\small{\it Simulation results showing plots of $-{\bf B}_\mathrm{rms}$ versus $ {\bf \omega_{rms}}$, for judiciously selected runs 1, 2, 3, 4, and 5. It is clear that a linear relationship holds exactly for the cases of $\mathrm{Pr_M}=1$. }}
       		 \includegraphics[height=2.5in]{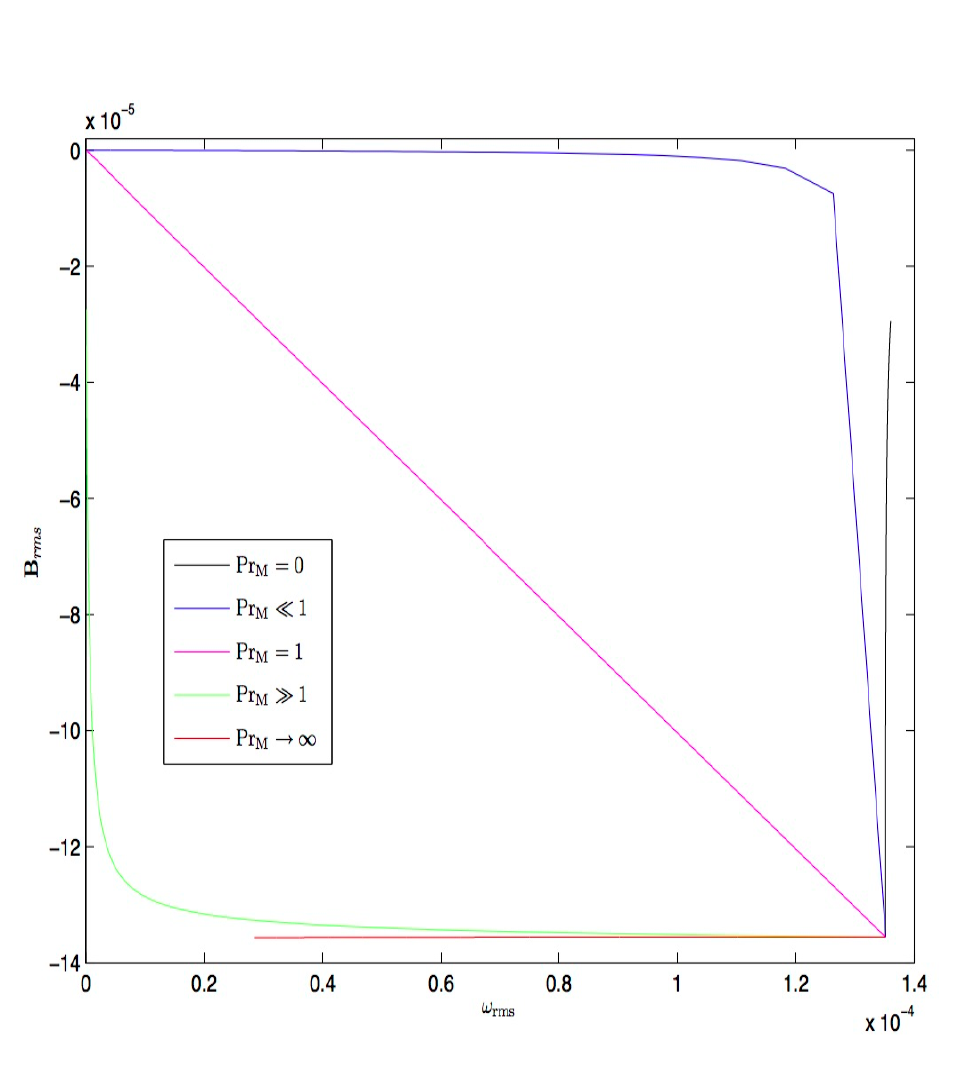}
       		\label{fig:1} 
\end{figure}

\begin{figure}[H]
 	\centering
\caption{\small{\it Simulation results showing the plots of $-{\bf B}_\mathrm{rms}$ versus $ \boldsymbol{\omega}_\mathrm{rms}$, for a range of runs $0 < Pr_M <\infty$. The linearity observed has been viewed as an indication of the existence of a possible analogy.}}
       		 \includegraphics[height=2.5in]{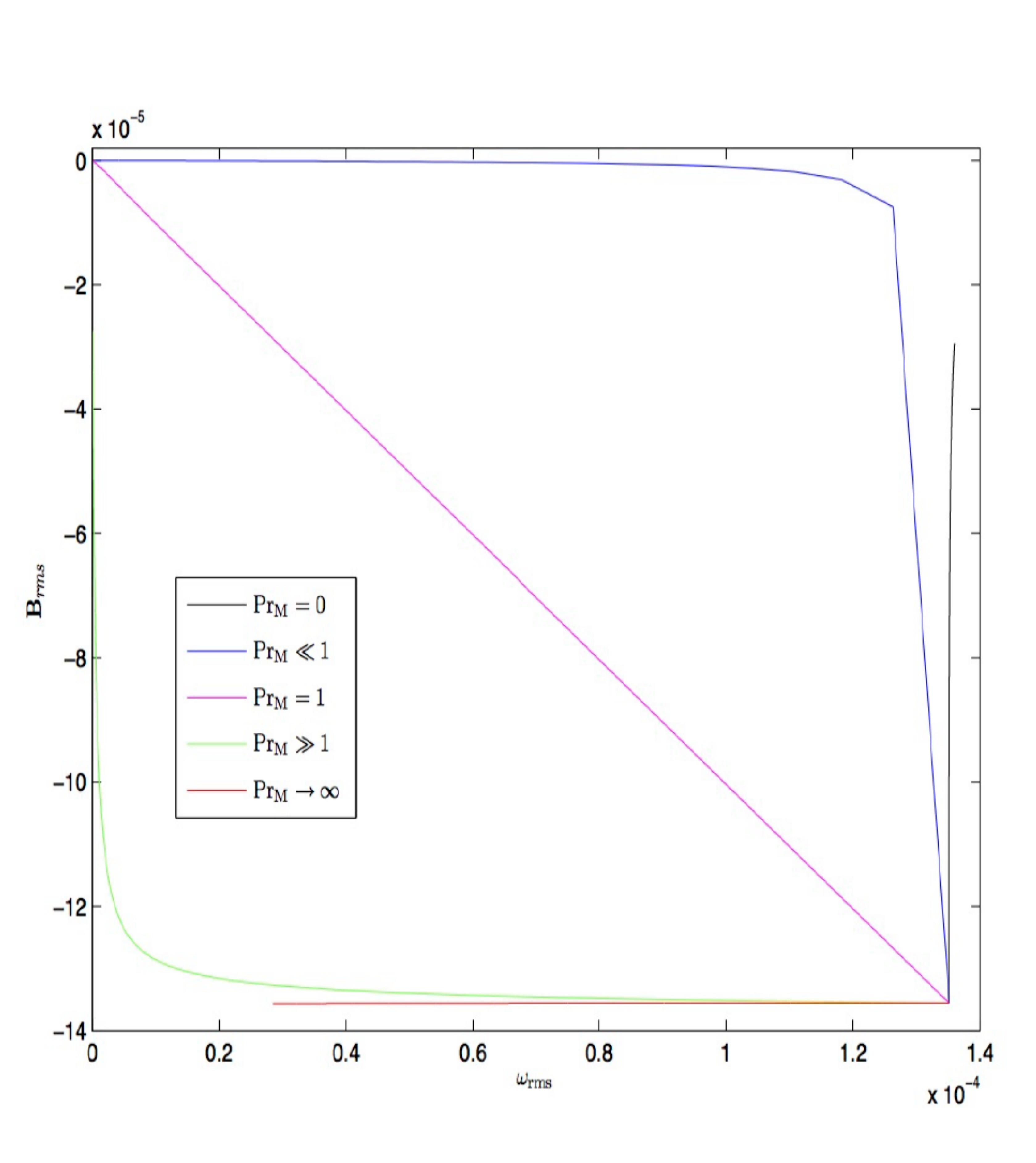}
     			 \label{fig:2}
  \end{figure}

In \cite{PB} simulation results which showed how magnetic and vorticity relate were presented. PENCIL CODE with a $32^3$ a periodic box of dimensions $2\pi\times2\pi\times2\pi$ was used. We revisit these results and compare ${\bf B}_{rms}$ to $\omega_{rms}$. The initial conditions were set to Gaussian noise of small amplitude and $\omega_{rms}$ and ${\bf B}_{rms}$ observed and compared. In this article, we only reconsider the results for several specific cases categorized by the following magnetic Prandtl number ($\mathrm{Pr_M}$) : $\mathrm{Pr_M} = 0$, $\mathrm{Pr_M} < 1$, $\mathrm{Pr_M} = 1$, $\mathrm{Pr_M} >1$ and $\mathrm{Pr_M} \rightarrow \infty$. Special attention is paid to cases where $\mathrm{Pr_M}\neq1$, and where the analogy between $\omega_{rms}$ and ${\bf B}_{rms}$ is tenuous at best. The following parameters (see Table \ref{tab:simpars2}) were used. \\\\
\begin{table}[H]
\vspace{-10mm}
	\centering
	{\renewcommand{\arraystretch}{1.2}%
		\begin{tabular}{|c|c|c|c|}
			\hline \hline
			Run & $\nu$ & $\eta$ & $\mathrm{Pr_M}$  \\
 			\hline 
			1& $0$ & $10^{-5}$ & $0$  \\
		 	2 & $10^{-3}$ & $10^{-1}$ & $10^{-2}$  \\
			3 & $10^{-3}$ & $10^{-3}$ & $1$  \\
			4 & $10^{-3}$ & $10^{-5}$ & $10^2$  \\
			5 & $10^{-5}$ & $\approxeq0$ & $\infty$  \\
			\hline \hline
		\end{tabular}}
	\caption{ \small{\it A summary of the simulation parameters for the runs used. We note that the values for $\nu$ and $\eta$ were arbitrarily chosen to achieve the desired $Pr_M$.}}
	\label{tab:simpars2} 
\end{table} \vspace{-3mm}Given these parameters, we obtained the results displayed in Figs. (\ref{fig:1} and \ref{fig:2}). In these figures we plot the values of ${\bf B}_{rms}$ against $\omega_{rms}$ for various $Pr_M$. It will be noticed that the extreme curves for $Pr_M=0$ and $Pr_M\rightarrow\infty$ are respectively vertical or horizontal and stop suddenly. These are due values of viscosity and diffusivity which affect the time step lengths resulting in the simulations running for a shorter "simulation time" than the others. Nevertheless, the curves for $0 < Pr_M < \infty$ suggest that the relationship between ${\bf B}_{rms}$ and $\omega_{rms}$ could be modelled  in a different way. 
~

  \section{Proposed ansatz}
\begin{figure}[H]
\centering
\caption{\small{\it $y=- x^{n}$, where $n=0.1,0.2,1,5,10$ and $0\leqq x\leqq1$}}
   \includegraphics[height=2.5in]{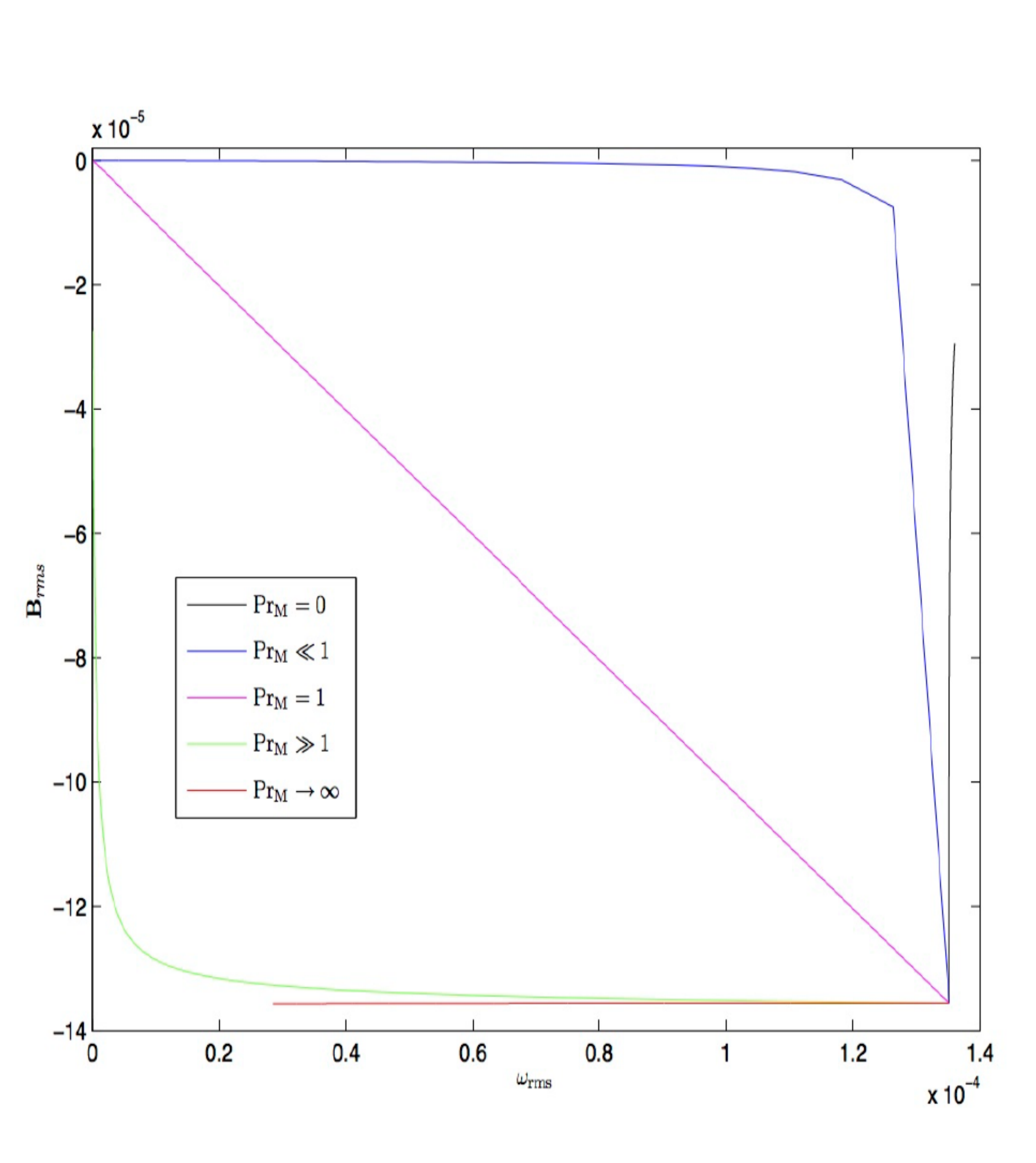}	   
	   \label{fig:3}
 \end{figure}

Figure (\ref{fig:2}) suggests that there exists some form of a power law relationship between ${\bf B}_{rms}$ and $\omega_{rms}$. This is the focus of this section and from which we propose a general analogue law. In particular, we examine $y=-x^{n}$ where $0 < n < \infty$; the plots of which are given in Fig.(\ref{fig:3}).
A comparison of Fig.(\ref{fig:2}) and Fig.(\ref{fig:3}) suggests that ${\bf B}_{rms}$ may be related to $\mathrm{\omega_{rms}}$ using an ansatz of the form ${\bf B}_{rms}\propto-\kappa(\omega_{rms})^{n}$, where $n=1/f(Pr_M)$ and where $\kappa$ is the constant of proportionality which may be given by the ratio of the initial value of  ${\bf B}_{rms}$ to the initial value of $\omega_{rms}$ ( i.e. ${{\bf B}_{rms}}_{0}/{\omega_{rms}}_{0}$), while $f(Pr_M)$ is a function of the Prandtl number. The simplest form of this relation is given by linear case of the function $f(Pr_M)$; in particular $f(Pr_M)=Pr_M$.\\\\
\section{Effective system}
We define an effective vorticity in ${\omega}_{eff}=\kappa(\omega_{rms})^{1/f({Pr_m})}$. This suggests that the system to be compared be:
\begin{eqnarray}
\label{one3}\frac{\partial{\bold{{\bf B}}}}{\partial{t}}&=&\nabla\times(\bold{v}\times{{\bf B}})=\eta\nabla^{2}{{\bf B}},\\
\label{one4}\frac{\partial{\bold{{\omega}_{eff}}}}{\partial{t}}&=&\nabla\times(\bold{u}\times{{\omega}_{eff}})+\eta\nabla^{2}{{\omega}_{eff}},
\end{eqnarray} 
where $\bold{u}$ can be recovered from the fact that $ \bold{u}=(\nabla\times)^{-1}({\omega}_{eff}/\kappa)^{Pr_M}$. $(\nabla\times)^{-1}$ is a vector inverse curl operator \cite{Sah,Saff}. We note that the Lorentz force term has been set to zero in the corresponding Navier-Stokes equations. $Pr_M = 1$, when $f(Pr_M)=Pr_M$, returns the system to the standard system that has been discussed extensively in literature.

It is important to reiterate there are infinitely many sets of $\nu$ and $\eta$ that gives the same Prandtl numbers. For example, $Pr_M={\nu/\eta}= 0.1$ may be given by (a) $\nu =100$ and $\eta =1000$, or (b) $\nu = 0.01$ and $\eta = 0.1$. These two fluids have two different viscosities and will not behave in the same manner, given that they may also have different magnetic Reynolds numbers. This suggests that we may not find an identical curve for a given $Pr_M$, but rather identical pattern. To find matching curves we would need to apply approximation theory\cite{Apx}. What is nevertheless certain is that the general "curvy" pattern should be expected, and this is what we need to draw our preposition.\\ \\ \hspace{10mm}

\section{Conclusion}
We have compared the curves obtained when the function $y=-x^{n}$  are plotted to the curves obtained when ${\bf B}_{rms}$ is plotted against $\omega_{rms}$ for various Prandtl numbers. We find similarities in patterns that suggest that the relationship between magnetic flux density and vorticity field may be modeled by a power law. Rather than comparing ${\bf B}_{rms}$ to $-\omega_{rms}$ propose that ${\bf B}_{rms}$ could be compared to ${\omega}_{eff} $ = $\kappa(\omega_{rms})^{1/f(Pr_M)}$, as a first step in the attempt to obtain a more realistic model. This allows for an analogous relationship that transcends $Pr_M\approxeq1$. It is conceivable that a better and possibly exact nonlinear relationship exists between ${\bf B}_{rms}$ and $\omega_{rms}$ for the simple case presented here where the Lorentz force is neglected or set to zero. The inclusion of the Lorentz force will alter the MHD behavior in dramatic ways, and is something that we will investigate in future\cite{BPTurb}.\\ \\ \\
\section{Acknowledgement}
Bob Osano acknowledges URC funding support administered by the University of Cape Town. Patrick W M Adams acknowledges funding support from the National Research Foundation (NRF) of South Africa, as well as funding from the University of Cape Town, both administered by the Postgraduate Funding Office (PGFO) of the University of Cape Town. \bibliography{references}

\appendix

\end{document}